\documentclass[prb,aps,twocolumn]{revtex4}
\usepackage{graphicx}
\usepackage{dcolumn}
\usepackage{bm}
\begin{document}
\title {Ultra-low noise field-effect transistor from multilayer graphene}
\author{Atindra Nath Pal\footnote[1]{electronic mail:atin@physics.iisc.ernet.in}}
\author{Arindam Ghosh}
\address{Department of Physics, Indian Institute of Science, Bangalore 560 012, India}

\begin{abstract}
We present low-frequency electrical resistance fluctuations, or noise, in graphene-based field-effect devices with varying number of layers. In
single-layer devices the noise magnitude decreases with increasing carrier density, which behaved oppositely in the devices with two or larger
number of layers accompanied by a suppression in noise magnitude by more than two orders in the latter case. This behavior can be explained from
the influence of external electric field on graphene band structure, and provides a simple transport-based route to isolate single-layer
graphene devices from those with multiple layers.
\end{abstract}


\maketitle

Graphene, single atomic layer of hexagonal carbon atoms, has drawn a lot of interest because of its unusual electronic
properties\cite{Novoselov}. In particular, extensive research on single and bi-layer graphene has led to significant improvement in both
material properties, as well as fundamental understanding, for nanoelectronic applications~\cite{Novoselov2,Rise of gfn,castro,Ohta,oostinga}.
Carrier mobilities as high as $1\times10^{4}$~cm$^{2}$~V$^{-1}$~S$^{-1}$ are now obtained on SiO$_2$ substrate, which is considerably enhanced
($\sim 2\times10^{5}$~cm$^{2}$~V$^{-1}$~S$^{-1}$) in suspended graphene~\cite{suspended_Kim}. Recently, bilayer
graphene~\cite{mccann1,mccann2,castroneto} has also emerged as a promising material in nanoelectronics in which a tunable band gap can be
induced by the application of perpendicular electric field \cite{oostinga,Atin} or chemical doping~\cite{castro,Ohta}. In contrast however, the
behavior of few-layer graphene devices with three or more atomics layers remains relatively unexplored. The significance of this issue is
immense in view of the difficulty in large scale production of single and bilayer graphene devices, with several chemical
methods~\cite{chem_method}, such as those involving reduction of graphite oxide~\cite{graphite_oxide}, routinely producing high-quality
multilayers of graphene. Uncertainties exist at the theoretical front too, where the band structure calculations in trilayer graphene within the
tight-binding framework in the presence of an external electric field yielded both opening of a gap~\cite{gap_multi} and semi-metallic
behavior~\cite{mccann_tri}. Although recent experiments~\cite{tri_expt} with double-gated trilayer devices support the latter, conventional
time-averaged characterization schemes, such evolution of carrier mobility with increasing layer number~\cite{intrinsic_disorder}, seems to be
inadequate in  understanding the overall behavior of gated multilayer graphene.

\begin{figure}
\begin{center}
\includegraphics [width=6.5cm,height=9cm]{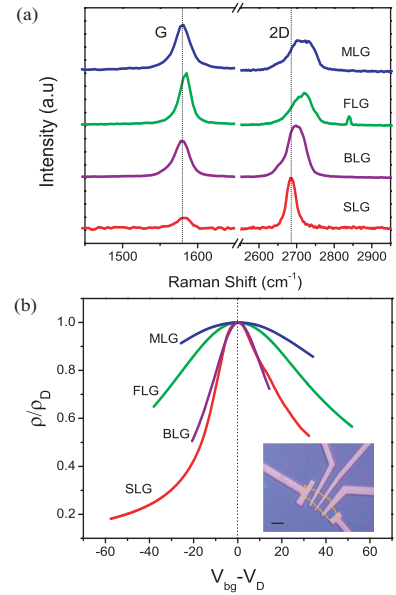}
\end{center}
\vspace{-.2cm}\caption{Color Online. (a) Raman spectra for SLG, BLG, FLG and MLG showing the characteristic G and 2D peaks. (b) Gate voltage
characteristics of graphene devices: for comparison, the ratio of resistivity ($\rho$) and the resistivity at the CNP ($\rho_D$) are plotted as
a function of ($V_{bg}- V_D$) at T = 100~K. The inset shows the optical micrograph and outline of a typical graphene device. The scale bar is $5
\mu$m. } \label{figure1}
\end{figure}

Being directly sensitive to the ability of an electronic device to screen external potential fluctuations, the low-frequency noise in electrical
transport has recently been shown to reflect the low-energy band structure in single and bilayer graphene devices~\cite{Atin,Avouris}. Although
the noise in both cases was found to originate from the disorder present in the SiO$_2$ substrate, primarily in the form of fluctuating charge
traps, the dependence of noise magnitude on the gate electric field was found to be opposite for single and bilayer graphene, and was attributed
to a field-induced gap formation in the latter. In the context of device application, however, both systems exhibited large noise magnitude
($\gamma_H \sim 1\times10^{-3}$, where $\gamma_H$ is the phenomenological Hooge parameter)~\cite{Atin,Avouris} similar to carbon nanotube FET
devices \cite{Avouris2,cnt2}, although some experiments~\cite{balandin} report lower values of $\gamma_H \sim 1\times10^{-4}$. As an emerging
new material, an investigation of noise performance of multilayer graphene devices is thus necessary for both application and fundamental
perspectives.

Here, we present a systematic study of low-frequency noise measurements in four categories of graphene devices: single layer graphene (SLG),
bilayer graphene (BLG), few-layer graphene (FLG) with $3-5$ atomic layers, and many-layer graphene (MLG) with greater than 5 layers in the
device. The noise magnitude at a given carrier density ($n$) was observed to be two orders of magnitude lower in the FLG and MLG devices in
comparison to that in SLG and BLG. Also, noise magnitude in SLG decreases with increasing $n$, while BLG, FLG and MLG devices show an increase
in noise on both side of the charge neutrality point (CNP), making noise in our case an excellent transport-based probe to identify single layer
graphene devices from the multilayered ones.

\begin{figure}
\begin{center}
\includegraphics [width=1\linewidth]{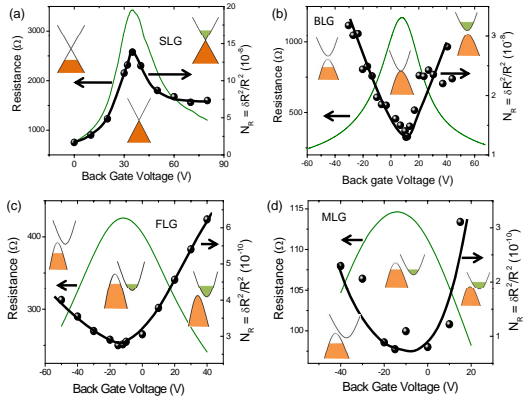}
\end{center}
\vspace{-.2cm}\caption{Color Online. The resistance and the normalized noise power spectral density ($N_R$) as functions of back gate voltages
are shown for: (a) SLG (b) BLG (c) FLG and (d) MLG devices at T = 100~K. The thick solid lines are guide to the eye. The insets in each figure
correspond to the bandstructure at particular voltages for corresponding graphene flake. } \label{figure2}
\end{figure}

Graphene flakes were prepared on $300$~nm SiO$_2$ on $n^{++}$ doped silicon substrate (the backgate) by micromechanical exfoliation of HOPG. All
flakes were characterized by Raman spectroscopy, and subsequent atomic force microscopy indicated the FLG and MLG devices in the present case to
consist of 3-4 and $\approx$ 14 layers, respectively. 40 nm gold contacts were defined using standard electron beam lithography technique.
Optical micrograph of a typical device is shown in the inset of Fig.~1b. All devices were prepared on identically processed substrate from the
same Si/SiO$_2$ wafer to keep the disorder level comparable. Fig.~1a shows the characteristic Raman spectra for different graphene flakes where
the intensity ratio of the G peak to the 2D peak can be seen to increase with increasing layer number \cite{ferrari}. Fig.~1b shows the gate
voltage characteristics of the devices. In all the cases, CNP was shifted to a finite gate voltage due to the intrinsic doping (see Fig.~2
also). Hence in order to compare the influence of gating we have plotted the ratio of resistivity to that at the CNP as a function of
$(V_{bg}-V_D)$, where $V_D$ is the back gate voltage at CNP. Fig.~1b clearly demonstrates the ambipolar transistor action in all devices,
although the effect of gating decreases with increasing layer numbers~\cite{Charge_FLG,ohta_multi}. Mobility of SLG, BLG, FLG and MLG devices
were calculated to be $1100$, $1160$, $2450$ and $1200$ cm$^{2}$~V$^{-1}$~S$^{-1}$ respectively.

Noise in the graphene devices were measured in low-frequency ac four-probe method. A carrier frequency of $777$~Hz was used to allow measurement
bandwidth of 256~Hz. Typical noise measurement involves digitization of the time-dependent output of the lockin amplifier, followed by
multistage decimation of the signal to eliminate effects of higher harmonic of the power line or other unwanted frequencies, and finally
estimation of the power spectral density. (See Ref [25] for details.) The excitation was below 50~$\mu$A to avoid heating and other
non-linearities, and verified by quadratic excitation dependence of voltage/current noise at a fixed resistance $R$. The background noise was
measured simultaneously, and subtracted from the total noise.

In all devices the normalized resistance noise power spectral density behaved as, $S_R(f) = \gamma_H R^2/nA_Gf^\alpha$, where $A_G$ is the area
of the flake between the voltage probes (not shown). The noise power spectral density was found to be proportional to $1/f^\alpha$, with
$\alpha$ ranging from $0.8-1.2$ (see Ref [11]). Here, instead of focusing on $\gamma_H$ or noise magnitude at a specific frequency, we compute
and analyze the total variance of resistance fluctuations $N_R = \langle\delta R^2\rangle/R^2 = (1/R^2)\int S_R(f)df$, where the integration is
carried out numerically over the experimental bandwidth. Fig.~2a-d shows the variation of $N_R$ and the corresponding average resistance as a
function of back gate voltage ($V_{bg}$) for SLG, BLG, FLG and MLG devices respectively. In case of SLG, noise decreases with increasing $n$ on
either side of the CNP (the Dirac point), which can be understood by better screening of potential fluctuations. But in all other cases, noise
magnitude behaves oppositely. In case of BLG, one can break the interlayer symmetry by applying a perpendicular electric field across the flake,
resulting in a gap between the conduction and valence band~\cite{mccann1,mccann2,castro}. Screening of the external potential fluctuations
weakens with increasing bandgap, leading to enhancement of noise at higher gate field~\cite{Atin,Avouris}. Moreover, the rate of change of noise
in SLG as a function of gate voltage was found to be larger than the other devices, indicating its different microscopic origin.

The qualitative similarity in the gate voltage dependence of noise in FLG and MLG to that of BLG naturally indicates a common underlying
physical mechanism. The band structure of gated three and four layer graphene has been recently carried out within the tight binding scheme with
Slonczewski-Wiess-McClure coupling parameters, taking screening into account in a self-consistent
manner~\cite{mccann_tri,gap_multi,gap_multi_solid state}. In the presence of a layer-symmetry breaking electric field on trilayers, however, the
theoretical calculations differ, although recent experimental results~\cite{tri_expt} suggest an enhancement in the overlap of conduction and
valence bands with increasing field. In our devices, however, the increase in noise in FLG and MLG with increasing $|n|$ suggests a reduction in
the density-of-state (DOS) at low energies, hence a field-induced pulling apart of the bands, which may eventually lead to a gap at the Fermi
energy~\cite{gap_multi}. (Note that in our case the thickness of the FLG is determined by atomic force microscopy, and hence uncertain between
three or four layers) The reason behind this difference is not clear, though a difference in stacking sequence in the FLG/MLG devices may be
envisaged. Indeed, it has been shown~\cite{gap_multi_solid state,Charge_FLG,New_J_phys} that while ABAB... Bernal stacking remains semi-metallic
under external field, a rhombohedral type (ABCA...) stacking displays opening of a gap under transverse electric field. Nevertheless, along with
time-averaged resistivity, noise seems to be a robust probe in this context, which can differentiate between such stacking modes in multilayer
graphene on thermodynamic (screening) grounds. Moreover, it is also important to note that, in these cases, noise measurements form an excellent
transport based technique, that immediately separates the SLG device from thicker ones without the need to go to the quantum Hall regime. It
also does not require Hall probes and hence can be used for FLG nanoribons.

\begin{figure}
\begin{center}
\includegraphics [width=6.5cm,height=5.5cm]{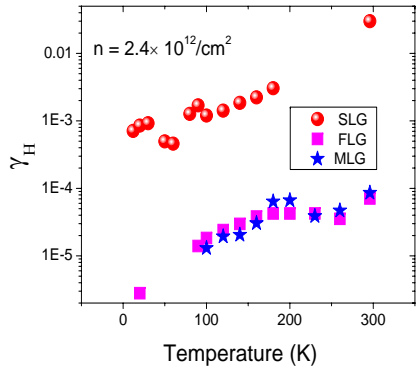}
\end{center}
\vspace{-.2cm}\caption{Color Online. (a) Temperature dependence of Hooge parameter for SLG, FLG and MLG devices. (b) Hooge parameter as a
function of graphene thickness for three different temperatures, $20$~K, $100$~K and $296$~K, far from the CNP (n = $2.4\times10^{12}$/cm$^2$).
} \label{figure3}
\end{figure}

To compare the noise performance in different devices, we have calculated the Hooge parameter $\gamma_H$ in SLG, FLG and MLG, at a particular
$n$ ($\approx 2.4\times10^{12}$~cm$^{-2}$) over a wide temperature ($T$) range. In Fig.~3, the exponential increase in $\gamma_H$ in all devices
with increasing $T$ could be readily understood in the framework of charging/discharging of the trap states at SiO$_2$ and graphene interface
which are known to be activated processes. Our analysis indicates the activation energies $\approx (17\pm1.5)$ meV, $(17\pm0.6)$ meV and
$(22\pm2)$ meV in SLG, FLG and MLG respectively, being expectedly similar since the devices were fabricated on the same Si/SiO$_2$ wafer.
However, the most intriguing aspect of Fig.~3 is nearly two orders of magnitude lower noise in FLG and MLG in comparison to SLG, that was
consistently observed in other similar devices as well. A simple understanding of this can be obtained by considering that at low energies
screening in multilayer graphene is primarily due to the parabolic bands (DOS $\sim m^*/\pi\hbar^2$), since the DOS tends to zero for the linear
bands at low energies. With increasing layer number the effective mass $m^*$ increases~\cite{tri_expt,effective_mass} , leading to increase in
the DOS, and hence reduction in the Thomas-Fermi screening length. Indeed, with $\gamma_H \sim 10^{-6} - 10^{-5}$ FLG forms a promising material
for low-noise nanoelectronic applications.

In summary, we have done a comperative study of low-frequency fluctuation in electrical resistance of various graphene based field effect
devices. The gate voltage characteristics of noise between multilayer graphene and SLG. A striking observation in this study is the extremely
low magnitude of noise in case of multilayer graphene, with Hooge parameter as low as $10^{-6}$ at low temperatures, making few/multilayer
graphene an attractive candidate for future nanoelectronics.

\textbf{Acknowledgement} We acknowledge the Department of Science and Technology (DST) for a funded project, and the Institute Nanoscience
Initiative, Indian Institute of Science, for infrastructural support. ANP thanks CSIR for financial support.

\end{document}